\documentclass[prd,12pt]{article}
\pdfoutput=1 

\usepackage{tikz}
\usepackage{amsmath,amssymb,graphicx,multirow,xspace,slashed,array,booktabs}
\usepackage[colorlinks=true,urlcolor=blue,anchorcolor=blue,citecolor=blue,filecolor=blue,linkcolor=blue,menucolor=blue,pagecolor=blue]{hyperref}
\usepackage[compress,numbers]{natbib}
\usepackage{subfigure, placeins}
\usepackage{booktabs}
\usepackage{blindtext, rotating}
\usepackage{afterpage}
\usepackage{enumitem}
\usepackage{ marvosym }
\usepackage{authblk} 
\usepackage{verbatim}
\usepackage{soul} 
\usepackage[normalem]{ulem}
\usepackage{pifont}

\makeatletter
	
	\@addtoreset{equation}{section}
\makeatother

\usepackage{feynmf}
\usepackage[compat=1.1.0]{tikz-feynman}

\newcommand{\bm}[1]{{\mbox{\boldmath $#1$}}}

\usepackage{floatrow}
\newfloatcommand{capbtabbox}{table}[][\FBwidth]

\usepackage[font=footnotesize,labelfont=bf]{caption}

\usepackage{lineno}

\allowdisplaybreaks

\long\def\symbolfootnote[#1]#2{\begingroup%
\def\thefootnote{\fnsymbol{footnote}}\footnote[#1]{#2}\endgroup}

\allowdisplaybreaks

\newcommand{\newc}{\newcommand}
\newc{\gsim}{\lower.7ex\hbox{$\;\stackrel{\textstyle>}{\sim}\;$}}
\newc{\lsim}{\lower.7ex\hbox{$\;\stackrel{\textstyle<}{\sim}\;$}}
\newc{\gev}{\,{\rm GeV}}
\newc{\mev}{\,{\rm MeV}}
\newc{\ev}{\,{\rm eV}}
\newc{\kev}{\,{\rm keV}}
\newc{\tev}{\,{\rm TeV}}
\newc{\MHT}{$H_T^{\text{miss}}$}
\newc{\MET}{$\slashed{E}_T$}
\newc{\MTT}{$M_{T2}$}

\def\dd{\mathrm{d}}

\def\ln{\mathop{\rm ln}}

\newc{\mz}{M_Z}
\newc{\mpl}{M_*}
\newc{\mw}{m_{\rm weak}}
\newc{\nr}[1]{N^c_R{}_{#1}}

\usepackage{feynmf}
\usepackage[compat=1.1.0]{tikz-feynman}

\def\dd{\mathrm{d}}
\def\Mpl{M_{\rm Pl}}
\newcommand{\beal}{\begin{equation}\begin{aligned}}
\newcommand{\enal}{\end{aligned}\end{equation}}

\def\beq{\begin{equation}}
\def\eeq{\end{equation}}
\newcommand{\bea}{\begin{eqnarray}\begin{aligned}}
\newcommand{\eea}{\end{aligned}\end{eqnarray}}
\def\bitem{\begin{itemize}}
\def\eitem{\end{itemize}}

\usepackage{url}
\usepackage{hyperref}

\begin{document}
\baselineskip 0.6cm

\begin{titlepage}

\vspace*{-0.5cm}

\thispagestyle{empty}

\begin{center}

\vskip 1cm

{\LARGE \bf Light Dark Photon Dark Matter \\[1ex] from Inflation}

\vskip 1cm

\vskip 1.0cm
{\large Yuichiro Nakai$^1$, Ryo Namba$^1$, and Ziwei Wang$^2$}
\vskip 1.0cm
{\it 
$^1$Tsung-Dao Lee Institute and School of Physics and Astronomy, \\Shanghai
Jiao Tong University, 800 Dongchuan Road, Shanghai, 200240, China \\
$^2$Department of Physics, McGill University, Montr\'eal, QC, H3A 2T8, Canada} \\
\vskip 1.0cm

\end{center}

\vskip 1cm

\begin{abstract}

We discuss the possibility of producing a light dark photon dark matter
through a coupling between the dark photon field and the inflaton.
The dark photon with a large wavelength is efficiently produced due to the inflaton motion during inflation
and becomes non-relativistic before the time of matter-radiation equality.
We compute the amount of production analytically.
The correct relic abundance is realized with a dark photon mass extending down to $10^{-21} \, \rm eV$.

\end{abstract}

\flushbottom

\end{titlepage}

\tableofcontents

\section{Introduction}

The nature of dark matter (DM) still remains unknown, while its existence has been established.
Although a weakly interacting massive particle (WIMP) is an intriguing possibility, there has been no evidence
which indicates its existence so far.
Given the situation, it is natural to keep an open mind on the identity of DM and search various other possibilities.
A spin-one dark photon, which arises from an abelian gauge group outside of the Standard Model,
gives one of the most popular alternatives to a WIMP DM in that dark photons are ubiquitous in physics beyond the Standard Model as well as string theory. 
A light dark photon DM does not require any symmetry to ensure its stability unlike
the case of a WIMP DM whose mass is around the electroweak scale.
Dark photons kinetically mixing with the ordinary photon have been actively searched over a wide range of their masses
(for a review of ongoing and future experiments, see ref.~\cite{Essig:2013lka}).

A dark photon DM with the correct relic abundance can be produced by a misalignment mechanism like axions \cite{Nelson:2011sf}
only if the dark photon field has finely-tuned non-minimal couplings to gravity \cite{Arias:2012az}.
Inflationary fluctuations can also produce the longitudinal mode of a dark photon DM through the mass term \cite{Graham:2015rva}.
The correct relic abundance is realized when the dark photon has a mass $m_{\gamma'} \gsim \mu{\rm eV}$.
A light dark photon whose mass is below $\mu$eV needs another mechanism.
Recently, this issue has been discussed by several authors \cite{Agrawal:2018vin,Dror:2018pdh,Co:2018lka}.
Their mechanisms, relying on a dark photon coupling to an axion-like field or a scalar field charged under a dark $U(1)$, can realize a dark photon DM with a wide range of masses extending down to $10^{-20} \, \rm eV$,
which reaches the fuzzy DM paradigm \cite{Hu:2000ke,Hui:2016ltb}.
The dark photon DM is produced with the assistance of another light scalar field coupling to the dark photon.
Ref.~\cite{Long:2019lwl} has discussed the production of a light dark photon DM from a network of cosmic strings.
Such cosmic strings arise from the same physics that makes the dark photon massive.

In this paper, we pursue the possibility of a light dark photon DM
without the assistance of any additional field other than the inflaton, which is by itself an essential ingredient for the standard cosmology.
Our dark photon field directly couples to the inflaton.
The authors of ref.~\cite{Bastero-Gil:2018uel} initiated such a possibility by introducing a coupling $\varphi F' \widetilde{F}'$
where $\varphi$ is the inflaton field and $F'$, $\widetilde{F}'$ are the field strength of the dark photon field $A'$ and its dual.
The dark photon DM is generated via tachyonic instability during inflation.
However, since this coupling is proportional to the wavenumber,
the dark photon field with a large wavelength is not efficiently produced and
the correct DM abundance could be achieved with a dark photon mass $m_{\gamma'} \gsim \mu{\rm eV}$,
which is not our target.
Instead, we consider a coupling of the dark photon field to the inflaton $I^2(\varphi)F'F'$
where $I$ is some function of $\varphi$.
The motion of $\varphi$ through this coupling breaks conformal invariance in the 4-D free Maxwell theory so that
the classical dark photon field is produced upon the horizon exit of quantum fluctuations during inflation.%
\footnote{Note that a free Maxwell field does not ``feel'' the horizon by itself due to the conformal invariance, and the presence of the coupling is essential. This fact is in sharp contrast to other matter contents whose free fields can be produced by coupling to gravity alone.}
This class of coupling has been actively studied in the context of primordial large-scale magnetogenesis \cite{Ratra:1991bn}, though its difficulty has also been evidenced \cite{Demozzi:2009fu, Barnaby:2012tk, Fujita:2012rb, Fujita:2013qxa, Fujita:2014sna, Ferreira:2013sqa, Ferreira:2014hma, Green:2015fss, BazrafshanMoghaddam:2017zgx} (see also \cite{Kobayashi:2014sga, Fujita:2016qab}).%
\footnote{The large-wavelength modes produced in the mechanism of this type can also source background and statistical anisotropies \cite{Kanno:2008gn, Watanabe:2009ct, Watanabe:2010fh, Gumrukcuoglu:2010yc, Bartolo:2012sd, Naruko:2014bxa}.}
Instead, we here show that the dark photon field with a large wavelength is efficiently produced with this coupling and
becomes non-relativistic before the time of matter-radiation equality.
The correct relic abundance is realized with a dark photon mass extending down to $10^{-21} \, \rm eV$.

The rest of the paper is organized as follows. In section~\ref{setup}, we present our simple coupled inflaton-dark photon system
to produce a light dark photon DM.
The equation of motion for each mode of the dark photon field is analytically solved in section~\ref{EOMsolutions}.
In section~\ref{DMabundance}, we show the final relic abundance of the light dark photon DM.
Section~\ref{discussion} is devoted to conclusions and discussions.
In Appendix~\ref{longitudinalcontribution}, we discuss the contribution of the longitudinal mode and demonstrate that it is subdominant to the transverse modes.

\section{The setup}\label{setup}

We here present our setup of a dark photon with non-zero mass that is coupled to the inflaton and our assumption of an inflationary background.
We then see that the dark photon field is decomposed into transverse and longitudinal modes around the isotropic background.
The equation of motion for each mode is derived.

\subsection{The inflaton-dark photon system}

We explore the production mechanism of a light dark photon DM during the inflationary epoch.
In order to break the conformal invariance of a free massless dark photon and to drive its production, we here investigate a simple mechanism that can induce a highly efficient production of light vector boson by a kinetic coupling to the inflaton $\varphi$. The relevant part of the action we are interested in is
\begin{equation}
S = \int \dd^4 x \sqrt{-g} \left[ \frac{\Mpl^2}{2} \, R 
- \frac{1}{2} \, \nabla_\mu \varphi \, \nabla^\mu \varphi - V(\varphi) 
- \frac{I^2(\varphi)}{4} \, F'_{\mu\nu} F'^{\mu\nu} - \frac{m_{\gamma'}^2}{2} \, A'_\mu A'^\mu \right] \; , \label{lagrangian}
\end{equation}
where we take the $(-+++)$ metric convention, $\Mpl$ is the reduced Planck mass, and $R$ and $\nabla_\mu$ are the Ricci scalar and covariant derivative, respectively, compatible with the metric $g_{\mu\nu}$. 
The function $V(\varphi)$ is the inflaton potential. The dark photon field $A'_\mu$ couples to the inflaton $\varphi$ through its kinetic term with a function of $\varphi$ while its field-strength tensor is $F'_{\mu\nu} \equiv \partial_\mu A'_\nu - \partial_\nu A'_\mu$. We have introduced a Stueckelberg/Proca mass $m_{\gamma'}$ for the dark photon, to identify it with the dark matter that is non-relativistic today. In principle, $m_{\gamma'}$ can depend on $\varphi$ as well, though we omit this dependence for our current attempt.
The crucial ingredient for our mechanism of the dark photon production is the modulation of the kinetic term, or equivalently of the dark $U(1)$ charge, by the inflaton field, captured by the function $I(\varphi)$.

During the inflationary epoch, we assume the following $1$-point vacuum expectation values (VEVs) of the inflaton and the dark photon field:%
\footnote{As we compute in Sec.~\ref{EOMsolutions}, the dark photon production in our mechanism results in a spectrum that is peaked at super-horizon scales. This implies that it can be viewed as a homogeneous (but not necessarily isotropic) ``background'' in the perspective of sub-horizon observers \cite{Bartolo:2012sd}. However, we ensure that the energy density of the produced dark photon is negligible to that of the inflaton, as discussed in \eqref{Friedmann} and below. Therefore our perturbative treatment of the dark photon field around the null value is self-consistent.
\label{foot:treatment}}
\begin{equation}
\langle \varphi \rangle \equiv \phi(t) \ne 0 \; , \qquad
\langle A'_\mu \rangle = 0 \; .
\end{equation}
The inflaton VEV is in general time-dependent.
Then the flat Friedmann-Lema\^{i}tre-Robertson-Walker (FLRW) metric is a consistent geometry for the background spacetime during inflation,
\begin{equation}
\dd s^2 = 
a^2(\tau) \left( - \dd\tau^2 + \delta_{ij} \, \dd x^i \dd x^j \right) \; ,
\end{equation}
where $\tau$ is the conformal time, related to the physical time $t$ through $\dd \tau = \dd t / a$, $a(\tau)$ is the scale factor, and $i=1,2,3$ is the spatial index. 
In the pure de Sitter background, $a$ can be related to $\tau$ as
$a = \frac{-1}{H \tau}$ with the physical Hubble parameter $H \equiv \partial_\tau a / a^2$ being constant.
The asymptotic past is $\tau \to -\infty$ while inflation proceeds to $\tau \to 0$.
For the inflationary scenario,
we assume a standard slow-roll inflation, which is an attractor solution provided the slow-roll conditions,
\begin{equation}
\begin{split}
\epsilon_V \equiv \frac{\Mpl^2}{2} \left( \frac{V_\varphi}{V} \right)^2 \ll 1 \; , \qquad
\eta_V \equiv \Mpl^2 \, \frac{V_{\varphi\varphi}}{V} \ll 1 \; ,
\end{split}
\end{equation}
are satisfied. Here, the subscript $\varphi$ of $V$ denotes derivative with respect to $\varphi$.

Along the inflationary trajectory, the time-dependent inflaton VEV $\phi(t)$ results in a time-evolving classical value of the dark photon coupling, $\langle I(\varphi) \rangle \equiv I(\phi)$.
For the sake of analytical control, let us take an ansatz on the time dependence of $I(\phi)$ during inflation.
We assume that this coupling changes for $a_i < a < a_{\rm end}$, where $a_i$ ($a_{\rm end}$) is the value of the scale factor where inflation starts (ends).%
\footnote{In a more general setup, one might cook up a nontrivial form of the coupling function such that $I(\phi)$ changes only for part of the inflationary period. In the spirit of minimal assumptions, however, we here identify the duration with the entire length of inflation.}
In principle, the coupling can be an arbitrary function of time, or the scale factor $a$. However, in order to simplify the analysis,
we postulate the following time dependence of the coupling during inflation:
\begin{equation}
\begin{split}
I (\phi) =  \left( \frac{a}{a_{\rm end}} \right)^n \approx \biggl( \frac{\tau_{\rm end}}{\tau} \biggr)^n \; ,
\label{formofI}
\end{split}
\end{equation}
for $a_i < a < a_{\rm end}$, with some real number $n$ and $\tau_{\rm end} =  -(H a_{\rm end})^{-1}$.
The last approximate equality is valid at the leading order in the slow-roll expansion.
We normalize $I$ such that its value reaches $1$ when inflation ends and stays constant thereafter, to recover the canonical kinetic term of the dark photon field.
With the slow-roll approximation, this time dependence of $I(\phi)$ can be induced by the form of the coupling
\begin{equation}
\begin{split}
I (\varphi) \approx {\rm e}^{-\frac{n}{\Mpl^{2}} \int_{\varphi_{\rm end}}^\varphi d \varphi' \frac{V(\varphi')}{V_\varphi(\varphi')}  }   \; ,
\label{formofi2}
\end{split}
\end{equation}
for an arbitrary $V(\varphi)$.
Thus, the function $I (\varphi)$ is determined by specifying the inflaton potential $V(\varphi)$.
For instance, in the case of a hilltop model \cite{Boubekeur:2005zm} of the type
$V (\varphi) = \Lambda^4 \left( 1 - \frac{\varphi}{\mu} + \cdots \right)$, 
where $\Lambda$ and $\mu$ are mass scales, the coupling function is given by
$I (\varphi) \propto e^{c\varphi}$ ($c = n \mu / \Mpl^2$ is a constant) during inflation.
For more sophisticated inflation models such as the Starobinsky model \cite{Starobinsky:1980te} and $\alpha$ attractors \cite{Kallosh:2013yoa}, 
$I$ is given by a more complicated function of $\varphi$ to attain the time dependence \eqref{formofI}.
While we assume \eqref{formofI} in the following discussions,
the rest of this section is independent of the explicit form of the dark photon coupling to the inflaton.

\subsection{Transverse and longitudinal modes}

Since we have assumed $\langle A'_\mu \rangle = 0$, the dark photon field $A'_\mu$ enters into the action only from the quadratic order. Thus, in order to consider the production due to the homogeneous motion of the inflaton, we only need to look at
the quadratic part of the action of $A'_\mu$, that is,
\begin{equation}
\begin{split}
S^{(2)}_{\gamma'} = \frac{1}{2} \int \dd\tau \, \dd^3x \left[ I^2(\phi) \left( F'_{0i} F'_{0i} - \frac{1}{2} \, F'_{ij} F'_{ij} \right)
- a^2 m_{\gamma'}^2 \left( - {A'_0}^2 + A'_i A'_i \right) \right] \; . 
\label{S2_A}
\end{split}
\end{equation}
It is evident that $A'_0$ enters without time derivatives, and thus a variation with respect to it provides a constraint equation. Thanks to the background rotational symmetry, it is convenient to decompose $A'_i$ into transverse modes ${A'_i}^T$ and a longitudinal mode $\chi$ as
\begin{equation}
A'_i = {A'_i}^T + \partial_i \chi \; , \qquad
\partial_i {A'_i}^T = 0 \; .
\label{TLdecomp}
\end{equation}
One then immediately sees that the constraint equation can be solved for $A_0'$ in terms only of $\chi$, that is,
\begin{equation}
A'_0 = \frac{- I^2}{- I^2 \partial^2 + a^2 m_{\gamma'}^2} \, \partial_\tau \partial_i A'_i 
= \frac{- I^2 \partial^2}{- I^2 \partial^2 + a^2 m_{\gamma'}^2} \, \partial_\tau \chi \; ,
\label{A0const}
\end{equation}
up to appropriate boundary conditions.
Plugging \eqref{TLdecomp} and \eqref{A0const} into the action \eqref{S2_A}, we observe that the action is split into the part of the transverse modes ${A'_i}^T$ and the part of the longitudinal mode $\chi$,
\begin{equation}
\begin{split}
S_{\gamma'}^{(2)} & = S_T + S_L \; , \\[1ex]
S_T & = \frac{1}{2} \int \dd\tau \, \dd^3x
\left[ I^2 \left( \partial_\tau {A'_i}^T \partial_\tau {A'_i}^T
- \partial_i {A'_j}^T \partial_i {A'_j}^T \right)
- a^2 m_{\gamma'}^2 {A'_i}^T {A'_i}^T  \right] \; , \\[1ex]
S_L
& = \frac{1}{2} \int \dd\tau \, \dd^3x  \, a^2 m_{\gamma'}^2 \left[ 
\partial_\tau \chi \left( \frac{- I^2 \partial^2}{- I^2 \partial^2 + a^2 m_{\gamma'}^2} \, \partial_\tau \chi \right)
- \partial_i \chi \, \partial_i \chi \right] \; ,
\end{split}
\label{quadaction}
\end{equation}
up to total derivatives.
We now derive the linearized equation of motion for each of the transverse and longitudinal modes from these actions.

\subsubsection{The transverse modes}

Let us first consider the transverse mode sector $S_T$. 
We canonically normalize the field ${A'_i}^T$ by defining $V_i \equiv I(\phi) \, {A'_i}^T$, where $I(\phi)$ is a classical quantity.
Noting that $I(\phi)$ depends only on time, we have
\begin{equation}
\begin{split}
S_T = \frac{1}{2} \int \dd\tau \, \dd^3x \left[ 
\partial_\tau V_i \, \partial_\tau V_i 
- V_i \left( - \partial^2 - \frac{\partial_\tau^2 I}{I} + \frac{a^2 m_{\gamma'}^2}{I^2} \right) V_i \right] \; ,
\label{S2_V}
\end{split}
\end{equation}
where $\partial^2 \equiv \partial_i \partial_i$.
We now decompose $V_i$ into Fourier and polarization modes as
\begin{equation}
V_i(\tau, \bm{x}) = \sum_{\sigma} \int \frac{\dd^3k}{(2\pi)^{3/2}} \,
{\rm e}^{i \bm{k} \cdot \bm{x}} \, e_i^\sigma( \hat{k}) \, \hat{V}_{\sigma , \bm{k}} (\tau) \; ,
\label{Ai_decomp}
\end{equation}
where $e_i^\sigma (\hat{k})$ are orthonormal polarization vectors perpendicular to the momentum direction $\hat{k} \equiv \bm{k} / \vert \bm{k} \vert$.
The reality condition of $V_i$ requires $\sum_\sigma e_i^{\sigma *} (\hat{k}) \hat{V}_{\sigma, \bm{k}}^\dagger = \sum_\sigma e_i^\sigma (-\hat{k}) \hat{V}_{\sigma, -\bm{k}}$. 
Then the action \eqref{S2_V} becomes
\begin{equation}
S_T = \frac{1}{2} \sum_{\sigma} \int \dd\tau \, \dd^3k \left[ 
\partial_\tau \hat{V}_{\sigma, \bm{k}}^\dagger \, \partial_\tau \hat{V}_{\sigma, \bm{k}} 
- \left( k^2 - \frac{\partial_\tau^2 I}{I} + \frac{a^2 m_{\gamma'}^2}{I^2} \right) \hat{V}_{\sigma, \bm{k}}^\dagger \hat{V}_{\sigma, \bm{k}} \right] \; .
\label{S2_V_2}
\end{equation}
Due to the nature of parity-invariant system, the mode functions of the two polarizations are identical, while the background isotropy guarantees the mode function depends only on the magnitude of the momentum, $k \equiv \vert \bm{k} \vert$.
Thus, we can decompose $\hat{V}_\sigma$ into the creation/annihilation operators as~%
\begin{equation}
\hat{V}_{\sigma, \bm{k}}(\tau) = V_k(\tau) \, \hat{a}_{\sigma , \bm{k}} + V_k^*(\tau) \, \hat{a}_{\sigma , -\bm{k}}^\dagger \; , \qquad
\left[ \hat{a}_{\sigma , \bm{k}} , \, \hat{a}_{\sigma' , \bm{k}'}^\dagger \right] = \delta_{\sigma\sigma'} \, \delta^{(3)} (\bm{k} - \bm{k}') \; ,
\label{V_modefun}
\end{equation}
and also find the equation of motion for the mode function,
\begin{equation}
\partial_\tau^2 V_k + \left( k^2 - \frac{\partial_\tau^2 I}{I} + \frac{a^2 m_{\gamma'}^2}{I^2} \right) V_k = 0 \; .
\label{EOM_V}
\end{equation}
In the next section (Sec.~\ref{EOMsolutions}), we solve this equation with the explicit form of $I(\tau)$ presented in \eqref{formofI}, 
to finally calculate the relic abundance of the dark photon DM.

\subsubsection{The longitudinal mode}

We now turn to the longitudinal mode sector $S_L$.
As in the previous case, we 
Fourier-transform the longitudinal mode $\chi$ as
\begin{equation}
\chi(\tau , \bm{x}) = \int \frac{\dd^3k}{(2\pi)^{3/2}} \, {\rm e}^{i \bm{k} \cdot \bm{x}} \, \hat\chi_{\bm{k}}(\tau) \; ,
\label{chi_decomp}
\end{equation}
with the property $\hat{\chi}_{\bm{k}}^\dagger = \hat{\chi}_{- \bm{k}}$ due to the reality condition.
Then, the longitudinal sector of the quadratic action \eqref{quadaction} becomes
\begin{equation}
S_L = \frac{1}{2} \int \dd\tau \, \dd^3k \, a^2 m_{\gamma'}^2 \left[ 
\frac{I^2 k^2}{I^2 k^2 + a^2 m_{\gamma'}^2} \, \partial_\tau \hat\chi_{\bm{k}}^\dagger \, \partial_\tau \hat\chi_{\bm{k}}
- k^2 \hat\chi_{\bm{k}}^\dagger \hat\chi_{\bm{k}} \right] \; .
\end{equation}
Canonically normalizing $\hat{\chi}_{\bm{k}}$ as
\begin{equation}
\hat\chi_{\bm{k}}(\tau) = \frac{\hat{X}_{\bm{k}}(\tau)}{z_k(\tau)} \; , \qquad
z_k(\tau) \equiv \frac{a m_{\gamma'} \, I k}{\sqrt{I^2 k^2 + a^2 m_{\gamma'}^2}} \; ,
\label{Xdef}
\end{equation}
we obtain
\begin{equation}
S_L = \frac{1}{2} \int \dd\tau \, \dd^3k \, \left[ 
\partial_\tau \hat{X}_{\bm{k}}^\dagger \, \partial_\tau \hat{X}_{\bm{k}}
- \left( k^2 - \frac{\partial_\tau^2 z_k}{z_k} + \frac{a^2 m_{\gamma'}^2}{I^2} \right) \hat{X}_{\bm{k}}^\dagger \hat{X}_{\bm{k}} \right] \; ,
\label{S2_L}
\end{equation}
up to total derivatives.
As in the case of the transverse modes,
the mode function depends only on the magnitude of the momentum, $k \equiv \vert \bm{k} \vert$.
We then decompose $\hat{X}_{\bm{k}}$ into the creation/annihilation operators,
\begin{equation}
\hat{X}_{\bm{k}}(\tau) = X_k(\tau) \, \hat{a}_{L , \bm{k}} + X_k^*(\tau) \, \hat{a}_{L , -\bm{k}}^\dagger \; , \qquad
\left[ \hat{a}_{L , \bm{k}} , \, \hat{a}_{L, \bm{k}'}^\dagger \right] = \delta^{(3)} ( \bm{k} - \bm{k}' ) \; .
\label{X_modefun}
\end{equation}
We find the equation of motion for the mode function $X_k$ as
\begin{equation}
\partial_\tau^2 X_k + \left( k^2 - \frac{\partial_\tau^2 z_k}{z_k} + \frac{a^2 m_{\gamma'}^2}{I^2} \right) X_k = 0 \; .
\label{EOM_X}
\end{equation}
This equation of motion will be also solved in the following section.

\section{Solutions of the EOMs}\label{EOMsolutions}

Let us now calculate the energy density of the dark photon field produced during inflation.
We express the energy density in terms of the mode functions and then solve the equation of motion for each case of the transverse and longitudinal modes.

The energy density of the dark photon can be obtained from the energy-momentum tensor involving the dark photon field
$T_{\mu\nu}^{A'} \equiv - 2 \, \delta S[A'] / \delta g^{\mu\nu}$, that is, in the conformal frame,
\begin{equation}
\begin{aligned}
\rho_{\gamma'} & \equiv - T^{A' \, 0}{}_0 
= 
\frac{1}{2 a^4} \left[
I^2 \left( F'_{0i} F'_{0i} + \frac{1}{2} \, F'_{ij} F'_{ij} \right)
+ a^2 m_{\gamma'}^2 \left( {A'_0}^2  + A'_i A'_i \right)
\right] \; .
\end{aligned}
\end{equation}
Recalling the transverse/longitudinal decomposition given in \eqref{TLdecomp}
and the constraint equation for $A'_0$ in \eqref{A0const},
the above expression of the energy density $\rho_{\gamma'}$ can be separated into the following three parts:
\begin{equation}
\begin{aligned}
\rho_{\gamma'} = \rho_{\gamma',T} + \rho_{\gamma',L} + \partial_i \mathfrak{J}_i \; ,
\end{aligned}
\end{equation}
where
\begin{equation}
\begin{aligned}
\rho_{\gamma',T} & = \frac{1}{2 a^4} \left[
I^2 \left( \partial_\tau {A'_i}^T \partial_\tau {A'_i}^T 
+ \partial_i {A'_j}^T \partial_i {A'_j}^T \right)
+ a^2 m_{\gamma'}^2 {A'_i}^T {A'_i}^T
\right] \; , \\[1ex]
\rho_{\gamma',L} & = \frac{m_{\gamma'}^2}{2 a^2} \left[ \partial_\tau \chi 
\left( \frac{-I^2 \partial^2}{-I^2 \partial^2 + a^2 m_{\gamma'}^2} \, \partial_\tau \chi \right)
+ \partial_i \chi \, \partial_i \chi
\right] \; . 
\end{aligned}
\end{equation}
That is, $\rho_{\gamma',T}$ only involves the transverse modes ${A'_i}^T$ while $\rho_{\gamma',L}$ the longitudinal mode $\chi$.
The part $\mathfrak{J}_i$ is also a function quadratic in ${A'_i}^T$ and $\chi$ (including cross terms), but enters into the energy density only through the form of divergence. 
In fact, for the vacuum average of the energy density $\langle \rho_{\gamma'} \rangle$, the contribution from $\partial_i \mathfrak{J}_i$ vanishes, 
reflecting the fact that a vacuum-averaged quantity on a homogeneous background is also homogeneous and thus divergence-free. 
We use the decompositions \eqref{Ai_decomp} and \eqref{V_modefun} for $\rho_{\gamma',T}$ and
\eqref{chi_decomp}, \eqref{Xdef} and \eqref{X_modefun} for $\rho_{\gamma',L}$.
Hence, we obtain
\begin{equation}
\langle \rho_{\gamma'} \rangle = \langle \rho_{\gamma',T} \rangle + \langle \rho_{\gamma',L} \rangle \; ,
\end{equation}
where
\begin{eqnarray}
&& \!\!\!\!\!\!\!\!\! \!\!\!\!\! \!\!\!\!\! \langle \rho_{\gamma',T} \rangle =
\frac{1}{a^4} \int \frac{\dd^3k}{(2\pi)^3}
\left[
I^2 \, \partial_\tau \left( \frac{V_k}{I} \right) \partial_\tau \left( \frac{V_k^*}{I} \right)
+ \left( k^2 + \frac{a^2 m_{\gamma'}^2}{I^2} \right) V_k V_k^*
\right] \; ,
\label{rhoT_vev} \\[3ex]
&& \!\!\!\!\!\!\!\!\! \!\!\!\!\! \!\!\!\!\! \langle \rho_{\gamma',L} \rangle =
\frac{1}{2 a^4} \int \frac{\dd^3k}{(2\pi)^3}
\left[
z_k^2 \,
\partial_\tau \left( \frac{X_k}{z_k} \right) \partial_\tau \left( \frac{X_k^*}{z_k} \right)
+ \left( k^2 + \frac{a^2 m_{\gamma'}^2}{I^2} \right) X_k X_k^*
\right] \; .
\label{rhoL_vev} \\ \nonumber
\end{eqnarray}
The remaining task is to compute $V_k$ and $X_k$ by solving their equations of motion. 
One noteworthy observation is that the forms of the quadratic action of the canonical transverse \eqref{S2_V_2} and the longitudinal \eqref{S2_L} modes, and therefore their respective equations of motion and energy densities, would be identical under the replacement $I \leftrightarrow z_k$, as can be seen in \eqref{EOM_V}, \eqref{EOM_X}, \eqref{rhoT_vev} and \eqref{rhoL_vev}.%
\footnote{The difference between \eqref{rhoT_vev} and \eqref{rhoL_vev} by a factor of $2$ is simply due to the difference in the number of degrees of freedom.}

In fact, as estimated in Appendix~\ref{longitudinalcontribution},
it turns out that the contribution of the longitudinal mode to the energy density is subdominant, compared to that of the transverse modes \eqref{rho_approx}, namely $\langle \rho_{\gamma'} \rangle \simeq \langle \rho_{\gamma',T} \rangle$.
We thus show the calculation of the transverse modes in detail below, leaving the consideration of the longitudinal to Appendix~\ref{longitudinalcontribution}.

All the expressions are exact up to this point, while the following calculations in this section are performed in the pure de Sitter background.
To proceed the computation analytically, we assume that the inflaton coupling to the dark photon $I(\phi)$ takes the form \eqref{formofI}. 
We here define $x \equiv - k \tau$ 
and assume 
$m_{\gamma'} / I \ll H$ during the period of inflation to avoid mass suppression of the production.
Then, the equation of motion for the transverse modes \eqref{EOM_V} is approximated as
\begin{equation}
\partial_x^2 V_k + \left[ 1 - \frac{n(n+1)}{x^2} \right] V_k \simeq 0 \; ,
\label{EOM_V_Dless_2}
\end{equation}
at the leading order in the slow-roll approximation.
The solution with the Bunch-Davies initial condition is given by
\begin{equation}
V_k (\tau) = i \, \frac{\sqrt{\pi}}{2} \sqrt{-\tau} \, H_{n + \frac{1}{2}}^{(1)} \left( - k \tau \right) \; ,
\label{Vlam-sol}
\end{equation}
up to an arbitrary constant phase, where $H_\nu^{(1)}$ is the Hankel function of the first kind. 
Plugging this solution into the expression of the energy density \eqref{rhoT_vev}, we obtain
\begin{equation}
\begin{aligned}
\langle \rho_{\gamma',T} \rangle
& =
\frac{H^4}{8\pi} \int_0^\infty \dd x \, x^2
\left[
x^2 \, \left\vert H_{n - \frac{1}{2}}^{(1)} \left( x \right) \right\vert^2
+ \left( x^2
+ \frac{m_{\gamma'}^2}{I^2 H^2} \right) 
\left\vert H_{n + \frac{1}{2}}^{(1)} \left( x \right) \right\vert^2
\right]
\; .
\end{aligned}
\label{rhogT_exact}
\end{equation}
Note that we keep the mass term in the above expression because the assumption $m_{\gamma'} / I \ll H$ does not \textit{a priori} guarantee $m_{\gamma'} / I \ll x H$ for super-horizon modes $x \ll 1$.
In addition, the energy density spectrum $P_{\gamma', T}$  is defined as
\begin{equation}
\begin{split}
\langle \rho_{\gamma' , T} \rangle \equiv \int  \dd \ln k \, P_{\gamma', T} (k)  \; .
\end{split}
\end{equation}
The small argument expansion of the Hankel function reads
\begin{equation}
\begin{split}
\Big\vert H_\nu^{(1)} (x) \Big\vert^2 \simeq \frac{\Gamma^2(\vert \nu \vert)}{\pi^2} \left( \frac{x}{2} \right)^{-2\vert\nu\vert} \; ,
\label{Hankel_expand}
\end{split}
\end{equation}
for $x \ll 1$ and $\nu \ne 0$.
Then, for super-horizon modes $- k \tau \ll 1$, the energy density spectrum is approximately given by
\begin{equation}
\begin{split}
P_{\gamma', T} (k) \simeq 
&\, \frac{H^4}{8 \pi^3}
\biggl[
2^{\left\vert 2 n - 1 \right\vert} \, \Gamma^2\left( \Big\vert n - \frac{1}{2} \Big\vert \right)
\left( - k \tau \right)^{5 - \vert 2n - 1 \vert} \\[1ex]
&+ 2^{\left\vert 2 n + 1 \right\vert} \, \Gamma^2\left( \Big\vert n + \frac{1}{2} \Big\vert \right)
\left( \left( - k \tau \right)^{5 - \vert 2n + 1 \vert} 
+ \frac{m_{\gamma'}^2}{I^2 H^2} \left( - k \tau \right)^{3 - \vert 2n + 1 \vert} \right) 
\biggr] \; .
\label{spectrum}
\end{split}
\end{equation}
Our interest is the generation mechanism of the dark photon with a sub-eV mass as the cold dark matter. In order for such a light dark photon to become non-relativistic before the time of matter-radiation equality, it is favored that most of the dark photon particles have small momenta.
This is achieved, under the current system, if the dark photon has a red spectrum,
that is, $P_{\gamma', T} (k)$ is a decreasing function of $k$. 
With this aim, recalling that \eqref{spectrum} is valid for $-k \tau \ll 1$ and $m_{\gamma'} / I \ll H$, we require
\begin{equation}
\vert n \vert > 2 \; .
\label{nrange}
\end{equation}
We are thus interested in this range of $n$.

To estimate the dark photon energy density, let us use the small-argument limit of the Hankel function in \eqref{Hankel_expand} to compute \eqref{rhogT_exact}
and take the hard-cutoff regularization by $k \lesssim -1/\tau$. 
Furthermore, in order to avoid the IR divergence, we set the smallest $k$ mode produced during inflation,
denoting it by $k_{\rm min} \equiv - \sqrt{n(n+1)} / \tau_i$.%
\footnote{The $\sqrt{n(n+1)}$ factor in the definition of $k_{\rm min}$ comes from the equation of motion \eqref{EOM_V_Dless_2}. The production occurs when a mode ``crosses the horizon,'' but this can happen for the light dark photon only through the change of $I$. Because of this, only the modes $- k \tau < \sqrt{n(n+1)}$ are produced, and the minimum among them is the one that experiences the crossing at the beginning of inflation, leading to $k_{\rm min} = - \sqrt{n(n+1)} / \tau_i$.}
Only the modes $k > k_{\rm min}$ get amplified.
We then have
\begin{equation}
\begin{aligned}
\langle \rho_{\gamma',T} \rangle & \simeq 
\frac{H^4}{8\pi^3} \int_{- k_{\rm min} \tau}^{{\cal O}(1)} \dd x \,
\biggl[\,
2^{\left\vert 2 n - 1 \right\vert} \, \Gamma^2\left( \Big\vert n - \frac{1}{2} \Big\vert \right)
x^{4 - \vert 2n-1 \vert} \\
&\qquad \qquad \qquad \qquad + 2^{\left\vert 2 n + 1 \right\vert} \, \Gamma^2\left( \Big\vert n + \frac{1}{2} \Big\vert \right)
\left( x^2 + \frac{m_{\gamma'}^2}{I^2 H^2} \right) 
x^{2 - \vert 2n+1 \vert}
\biggr]
\; .
\end{aligned}
\end{equation}
Recall the time dependence of $I$ as in \eqref{formofI}, i.e.~$I = (a / a_{\rm end})^n = (\tau_{\rm end} / \tau)^{n}$, where $\tau_{\rm end}$ is the conformal time when inflation ends.
Since $-k_{\rm min} \tau \ll 1$ and we have assumed $m_{\gamma'} / I \ll H$ for the period of our interest,
the mass term contribution turns out to be negligible after the integration.
Hence, evaluated at time $\tau = \tau_{\rm end}$, we get
\begin{equation}
\begin{split}
\langle \rho_{\gamma',T} \rangle_{\tau = \tau_{\rm end}} \simeq \frac{H^4}{8\pi^3} \times
\frac{2^{2 \vert n \vert} \, \Gamma^2\left( \vert n \vert + \frac{1}{2} \right)}{\vert n \vert - 2} \,
\frac{1}{( - k_{\rm min} \tau_{\rm end} )^{2 \vert n \vert - 4}} \; ,
\end{split}
\label{rho_approx}
\end{equation}
valid for $\vert n \vert > 2$.
Note $(- k_{\rm min} \tau_{\rm end})^{-1} \sim {\rm e}^{N}$, where
\begin{equation}
\begin{split}
N \equiv \ln \frac{a_{\rm end}}{a_i}
\; ,
\end{split}
\end{equation}
is the number of e-folds during inflation.
The produced modes are limited such that $-n/\tau_i \lesssim k \lesssim -n/\tau_{\rm end}$,
and the comoving wavenumber at which the spectrum is peaked is given by
\begin{equation}
\begin{split}
k_{\rm peak} \approx k_{\rm min} = \frac{- \sqrt{n(n+1)}}{\tau_i} \; .
\end{split}
\end{equation}
As noted earlier in this section as well as in Appendix \ref{longitudinalcontribution}, the contribution from the longitudinal mode is negligible, and the energy density is dominated by the transverse mode contribution.
Therefore, the estimate of \eqref{rho_approx} gives the total energy density of the dark photon at the end of inflation
$\langle \rho_{\gamma'} \rangle_{\tau = \tau_{\rm end}}$.
The present values of $\langle \rho_{\gamma'} \rangle$ and the physical peak momentum will depend on the cosmological history, which we will discuss in the next section.

Produced dark photons can back-react to the evolution of the homogeneous background during inflation, which puts the upper limit for the amount of dark photons.
Taking the vacuum average of the inflaton EOM, one finds that the homogeneous mode $\phi$ receives a source/drain term due to the produced dark photons, denoted by ${\cal S}_\phi$.
In order for ${\cal S}_\phi$ not to significantly back-react and to be treated only perturbatively, 
the condition is given by
\begin{equation}
\begin{split}
\vert {\cal S}_\phi \vert \ll 3 H \vert \dot\phi \vert \; ,
\qquad  {\cal S}_\phi  \equiv - \frac{I I_{\varphi}}{2} \, \left\langle F_{\mu\nu} F^{\mu\nu} \right\rangle \; ,
\label{backreaction}
\end{split}
\end{equation}
where dot denotes derivative with respect to the physical time $t$.
As discussed above, the energy density of produced dark photons is dominated by the transverse mode contribution.
Using the Fourier decomposition \eqref{Ai_decomp} and the solution \eqref{Vlam-sol},
we find, integrating momenta for $k_{\rm min} < k \lesssim - n / \tau$,
\begin{equation}
\begin{split}
\vert {\cal S}_\phi \vert & 
\simeq \frac{H^4}{4\pi^3} \, \frac{I_\varphi}{I} \times
\frac{2^{2 \vert n \vert} \,
\Gamma^2\left( \vert n \vert + \frac{1}{2} \right)}{\vert n \vert - 2} \,
\frac{1}{( - k_{\rm min} \tau )^{2 \vert n \vert -4}} \; ,
\label{backreaction2}
\end{split}
\end{equation}
for $\vert n \vert > 2$ at a given time $\tau$. Here, $
-k_{\rm min} \tau \sim {\cal O}(\tau / \tau_i) \ll 1$ and
we have used the small argument expansion of the Hankel function \eqref{Hankel_expand}.
Another backreaction condition comes from the Friedmann equation,
\begin{equation}
\begin{split}
\langle \rho_{\gamma'} \rangle \ll 3 \Mpl^2 H^2 \; .
\label{Friedmann}
\end{split}
\end{equation}
Comparing the conditions \eqref{backreaction} and \eqref{Friedmann},
\begin{equation}
\frac{\vert {\cal S}_\phi \vert}{3 H \vert \dot\phi \vert}
= \frac{I \vert \dot I \vert}{6 H \dot\phi^2} \, \big\vert \langle F^2 \rangle \big\vert
\simeq \frac{\vert n \vert}{2 \epsilon_V} \, \frac{\langle \rho_{\gamma'} \rangle}{3 \Mpl^2 H^2} \; ,
\end{equation}
we see the former condition \eqref{backreaction} is more stringent than the latter \eqref{Friedmann} by the slow-roll parameter $\epsilon_V^{-1}$.
The backreaction condition thus amounts to, using \eqref{backreaction} and \eqref{backreaction2},
\begin{equation}
\begin{split}
\frac{1}{-k_{\rm min} \tau} \ll
\left[
\frac{3 \pi \left( \vert n \vert - 2 \right)}{2^{2 \vert n \vert} \vert n \vert \, \Gamma^2 \left( \vert n \vert + \frac{1}{2} \right) P_\zeta}
\right]^{\frac{1}{2\vert n \vert - 4}} \; ,
\end{split}
\label{cond_backreaction}
\end{equation}
where the scalar power spectrum at the pivot scale is given by
\begin{equation}
\begin{split}
P_\zeta \, \vert_{k = k_*} \cong \left( \frac{H^2}{2\pi \dot\phi} \right)^2 \simeq \frac{H^2}{8 \pi^2 \epsilon_V \Mpl^2} 
\simeq 2.1 \times 10^{-9} \; .
\end{split}
\end{equation}
The physical meaning of the above conditions can be understood as follows: the energy that is used to produce the dark photon particles is transferred from the motion of the inflaton condensate. The conservation law requires the produced energy to be smaller than the producing one, resulting in the condition \eqref{Friedmann}. The act of the inflaton on the dark photon comes with a counteract by the dark photon. The produced dark photons disturb the motion of the inflaton condensate. If this disturbance were significant, the inflationary dynamics would be disrupted, which would at least invalidate our calculation. The condition \eqref{backreaction} is to avoid such an outcome. 
The latter condition turns out more stringent, and thus \eqref{cond_backreaction} ensures negligible amount of both the mentioned effects on the background dynamics.
The region of the parameter space excluded by this condition will be shown in the next section.

\section{The relic abundance}\label{DMabundance}

After the end of inflation at $a = a_{\rm end}$, reheating proceeds 
until $a = a_{\rm reh} \ge a_{\rm end}$. 
We assume the standard scenario that the effective equation of state after inflation until the completion of reheating is $w = 0$
and that all the energy is transferred to the radiation component.
The produced dark photon with a comoving wavenumber $k$ has the physical momentum,
\begin{equation}
q(t) = \frac{k}{a(t)} \; , \label{q-scaling}
\end{equation}
as a function of time $t$.
Since the comoving wavenumber at which the spectrum is peaked is $k_{\rm peak} \approx k_{\rm min}$,
the peak momentum at $t = t_{\rm reh}$ is given by
\begin{equation}
\begin{split}
q_{\rm peak} (t_{\rm reh}) = \frac{k_{\rm peak}}{a(t_{\rm reh})}
= - k_{\rm min} \tau_{\rm end} 
\left( \frac{\rho(t_{\rm reh})}{\rho(t_{\rm end})} \right)^{{1}/{3}} 
\; ,
\label{qpeak}
\end{split}
\end{equation}
where $-k_{\rm min} \tau_{\rm end}$ is the factor appearing in \eqref{rho_approx}.
Let us consider the case where the momentum satisfies $q_{\rm peak} (t_{\rm reh}) \gg m_{\gamma'}$,
which also prevents the mass suppression of the dark photon production. (The other scenario is considered later, eq.~\eqref{cond_nonrelat} and below.)
The dark photon just after 
the end of inflation is relativistic and its energy density
$\langle \rho_{\gamma'} \rangle$ decreases as $a^{-4}$,
but its momentum is red-shifted and finally it becomes non-relativistic at some time $t=t_{\rm NR}$ where
\begin{equation}
q_{\rm peak} (t_{\rm NR}) = m_{\gamma'} \; .
\label{q-nonrel}
\end{equation}
From the above equations, the temperature $T_{\rm NR}$ at $t=t_{\rm NR}$ is given by
\begin{equation}
\begin{split}
T_{\rm NR} &= m_{\gamma'} \, \frac{T_{\rm reh}}{q_{\rm peak} (t_{\rm reh})} \\
&= \left( \frac{30}{\pi^2 g_*(t_{\rm reh})} \right)^{1/4} \,
\frac{m_{\gamma'}}{H}
\left( \frac{\rho(t_{\rm end})}{\rho(t_{\rm reh})} \right)^{{1}/{12}} \,
\frac{\rho^{1/4}(t_{\rm end})}{- k_{\rm min} \tau_{\rm end}} \; .
\label{TNR_2}
\end{split}
\end{equation}
Here, $T_{\rm reh}$ is the reheating temperature and $g_*(t_{\rm reh})$ is the number of relativistic degrees of freedom at
$t = t_{\rm reh}$. 
Let us remind that $(- k_{\rm min} \tau_{\rm end})^{-1} \sim {\rm e}^{N}$ is a large factor.
After the dark photon becomes non-relativistic, the energy density decays as $\langle \rho_{\gamma'} \rangle \propto a^{-3}$.

The dark photon must be non-relativistic at the time of matter-radiation equality, $t=t_{\rm eq}$,
which satisfies $\rho_m(t_{\rm eq}) = \rho_r (t_{\rm eq})$ with $\rho_m$ being the matter density.
The temperature at $t=t_{\rm eq}$ can be determined by 
\begin{equation}
\begin{split}
T_{\rm eq} = \frac{90}{\pi^2 g_*(t_0)} \, \Omega_m \, \frac{\Mpl^2 H_0^2}{T_0^3} \; ,
\label{Teq}
\end{split}
\end{equation}
where $H_0$, $T_0$ and $g_*(t_{0})$ are the Hubble parameter, the (CMB) temperature
and the number of relativistic degrees of freedom at present, respectively.
The present fractional matter density is denoted by $\Omega_m \equiv \rho_m(t_0) / \rho(t_0)$ as usual.
From \eqref{TNR_2} and \eqref{Teq}, imposing $T_{\rm NR} > T_{\rm eq}$, we obtain the constraint,
\begin{equation}
\begin{split}
\frac{T_{\rm NR}}{T_{\rm eq}}
= \frac{\left( \pi^2 / 30 \right)^{2/3}}{\sqrt{3} \, \Omega_m} \,
\frac{g_*(t_0)}{g_*^{1/3}(t_{\rm reh})} \,
\frac{m_{\gamma'}}{\rho^{1/6}(t_{\rm end}) \, T_{\rm reh}^{1/3}} \, 
\frac{T_0^3}{\Mpl H_0^2} \,
\frac{1}{-k_{\rm min} \tau_{\rm end}}
\,\,> 1 \; .
\end{split}
\label{cond_nonrel}
\end{equation}
Another constraint comes from the masslessness of the dark photon during the production period.
For the approximated equation of motion \eqref{EOM_V_Dless_2} to be valid, $m^2_{\gamma'} / I^2 < |n(n+1)| H^2$ has to be satisfied. 
For $n < -2$, $I$ is a decreasing function of $a$ and
hence it is sufficient to impose the constraint at the end of inflation,
$t=t_{\rm end}$; for $n>2$, $I$ is an increasing function and
it has to be imposed at the starting time $t=t_i$. Thus the constraint translates to
\begin{equation}
\begin{split}
m_{\gamma'} < 
\frac{\rho^{1/2}(t_{\rm end})}{3^{1/2} \Mpl}
\times
\begin{cases}
\sqrt{\vert n(n+1) \vert} \; , \quad
& n < -2 \; , \vspace{1mm} \\
\sqrt{\vert n(n+1) \vert} \, {\rm e}^{- n N} \; , \quad
& n > 2 \; .
\end{cases}
\label{cond_masslessness}
\end{split}
\end{equation}
We will take into account of these two conditions when we discuss the parameter space to give the correct relic abundance of
the dark photon dark matter.

We now evaluate the present density of the dark photon dark matter.
The present energy density evolves from the end of inflation as
\begin{equation}
\langle\rho_{\gamma'} \rangle_{t=t_0} = \langle \rho_{\gamma'} \rangle_{t=t_{\rm end}}
\left( \frac{a_{\rm end}}{a (t_{\rm NR})} \right)^4 \left( \frac{a (t_{\rm NR})}{a_0} \right)^3 \; ,
\end{equation}
where $a_0$ is the present value of the scale factor.
The ratios of the scale factor above are given by
\begin{equation}
\begin{aligned}
\frac{a_0}{a (t_{\rm NR})}
= \frac{1}{2^{1/4}} \left( \frac{\pi^2}{30} \right)^{1/3}
\frac{g_*^{1/12}(t_{\rm eq}) \, g_*^{1/4}(t_{\rm NR})}{\rho^{1/3} (t_0) \, \Omega_m^{1/3}} \,
T_{\rm eq}^{1/3} \, T_{\rm NR} \; ,
\end{aligned}
\end{equation}
and
\begin{equation}
\begin{aligned}
\frac{a (t_{\rm NR})}{a_{\rm end}}
= \left( \frac{g_*(t_{\rm reh})}{g_*(t_{\rm NR})} \right)^{1/4}
\frac{T_{\rm reh}}{T_{\rm NR}}
\left( \frac{\rho(t_{\rm end})}{\rho(t_{\rm reh})} \right)^{1/3} \; .
\end{aligned}
\end{equation}
Combining the above expressions with \eqref{rho_approx},
we obtain the present fractional density of the dark photon as
\begin{equation}
\begin{split}
\Omega_{\gamma'} & \equiv \frac{\langle\rho_{\gamma'} \rangle_{t=t_0}}{\rho (t_0)}
\\[1ex]
& = 
\frac{7.609 \times 10^{-28}}{h^2} \,
\frac{2^{2 \vert n \vert - 4} \Gamma^2\left( \vert n \vert + \frac{1}{2} \right)}{\vert n \vert - 2}
\frac{g_*(t_0)}{\left( - k_{\rm min} \tau_{\rm end} \right)^{2 \vert n \vert - 3}}
\\ & \qquad
\times
\left( \frac{m_{\gamma'}}{10^{-15} \, {\rm eV}} \right)
\left( \frac{T_{\rm reh}}{10^{14} \, {\rm GeV}} \right)
\left( \frac{\rho^{1/4}(t_{\rm end})}{10^{16} \, {\rm GeV}} \right)^2
\; ,
\end{split}
\label{relic_relativistic}
\end{equation}
for $\vert n \vert >2$.
Here, we have used the numerical values, 
$T_0 = 2.725 \, {\rm K} = 2.348 \times 10^{-13} \, {\rm GeV}$,
$H_0 = 2.133 \times 10^{-42} \, h \, {\rm GeV}$ and
$\Mpl = 2.435 \times 10^{18} \, {\rm GeV}$
and assumed that the time $t_{\rm NR}$ is late enough such that the numbers of relativistic degrees of freedom is equal to that at the equality, i.e.~$g_*(t_{\rm NR}) = g_*(t_{\rm eq})$.
Let us emphasize that the small numerical factor in \eqref{relic_relativistic} is compensate by the huge factor $1/\left( - k_{\rm min} \tau_{\rm end} \right)^{2 \vert n \vert - 3}$ and thus the production mechanism we discussed in the previous section is crucial to achieve the correct relic abundance.

While we have so far assumed that the dark photon is relativistic during the production, it is possible that produced dark photons become non-relativistic before the end of inflation.
Since we are interested in the dark photon spectrum that is red,
this situation happens when
\begin{equation}
\begin{split}
\frac{k_{\rm peak}}{a_{\rm end}} < {m_{\gamma'}} \; .
\label{cond_nonrelat}
\end{split}
\end{equation}
We still respect the masslessness condition \eqref{cond_masslessness},
which is necessary to ensure the validity of our calculation of the transverse modes.
However, in the case of $n > 2$, the above condition \eqref{cond_nonrelat} is not compatible with the masslessness condition.
Thus, the current consideration applies only to the case of $n < -2$.
Combining \eqref{cond_nonrelat} and \eqref{cond_masslessness}, we find
\begin{equation}
\begin{split}
\frac{\rho^{1/2}(t_{\rm end})}{\sqrt{3} \, \Mpl} \sqrt{n(n+1)} \, {\rm e}^{- N}
< m_{\gamma'} <
\frac{\rho^{1/2}(t_{\rm end})}{\sqrt{3} \, \Mpl}
\sqrt{\vert n(n+1) \vert} \; , \quad
n < -2 \; .
\label{mrange_nonrel}
\end{split}
\end{equation}
In this case, as soon as inflation ends, the dark photon behaves as a non-relativistic matter,
and the estimate of its current energy density becomes,
\begin{equation}
\begin{split}
\langle\rho_{\gamma'} \rangle_{t=t_0} & = \langle \rho_{\gamma', T} \rangle_{t=t_{\rm end}} \left( \frac{a_{\rm end}}{a_0} \right)^3 
\; ,
\end{split}
\end{equation}
where $\langle \rho_{\gamma', T} \rangle_{t=t_{\rm end}}$ is given in \eqref{rho_approx}.
Hence the present fractional density in this scenario is
\begin{equation}
\begin{split}
\Omega_{\gamma'} & \equiv \frac{\langle\rho_{\gamma'} \rangle_{t=t_0}}{\rho_0} \\ & 
= \frac{1.804 \times 10^{10}}{h^2} \, 
\frac{2^{2 \left( \vert n \vert - 2 \right)} \, \Gamma^2\left( \vert n \vert + \frac{1}{2} \right)}{\vert n \vert - 2} \,
\frac{g_*^{1/4}(t_{\rm reh}) \, g_*(t_0)}{g_*^{1/4}(t_{\rm eq})} \,
\frac{1}{\left( - k_{\rm min} \tau_{\rm end} \right)^{2 \vert n \vert - 4}} \\
& \qquad \times \left( \frac{T_{\rm reh}}{10^{14} \, {\rm GeV}} \right) \,
\left( \frac{\rho_{\rm end}^{1/4}}{10^{16} \, {\rm GeV}} \right)^4
\; .
\end{split}
\label{relic_nonrel}
\end{equation}
This result is valid for the values of $m_{\gamma'}$ in the range of \eqref{mrange_nonrel}.
Note that the expressions \eqref{relic_relativistic} and \eqref{relic_nonrel} neatly coincide each other where they should, i.e.~at the point where the left inequality in \eqref{mrange_nonrel} saturates, which is a non-trivial cross-check of our calculation.
Due to the large numerical factor \eqref{relic_nonrel}, the reheating temperature and the energy scale of inflation need to be small for this scenario to give the correct abundance, and the value of $n$ should be close to $n=-2$.

\begin{figure}
\resizebox{0.46\columnwidth}{!}{
\includegraphics[scale=0.8]{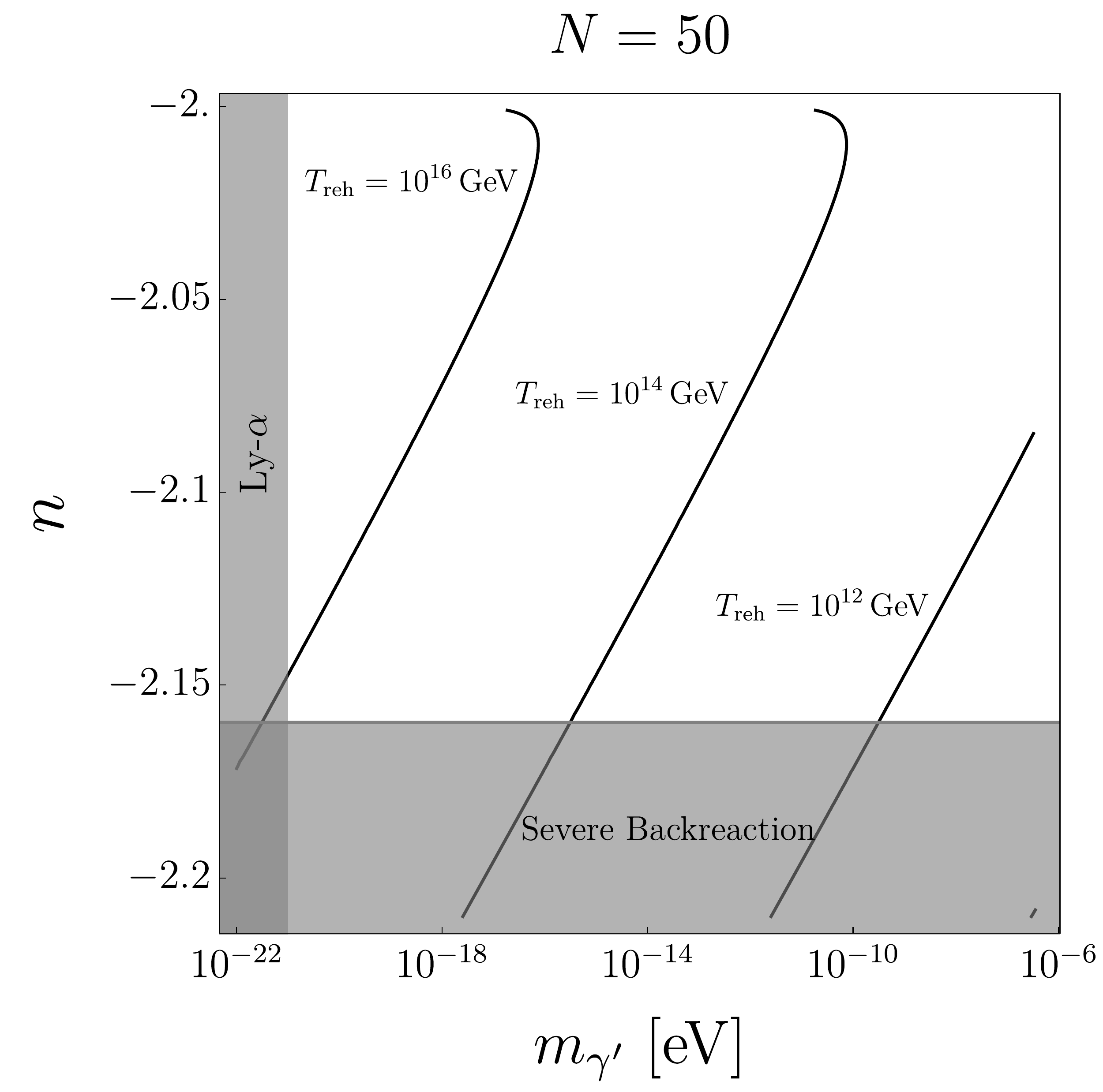}
}
\hspace{1cm}
\resizebox{0.46\columnwidth}{!}{
\includegraphics[scale=0.8]{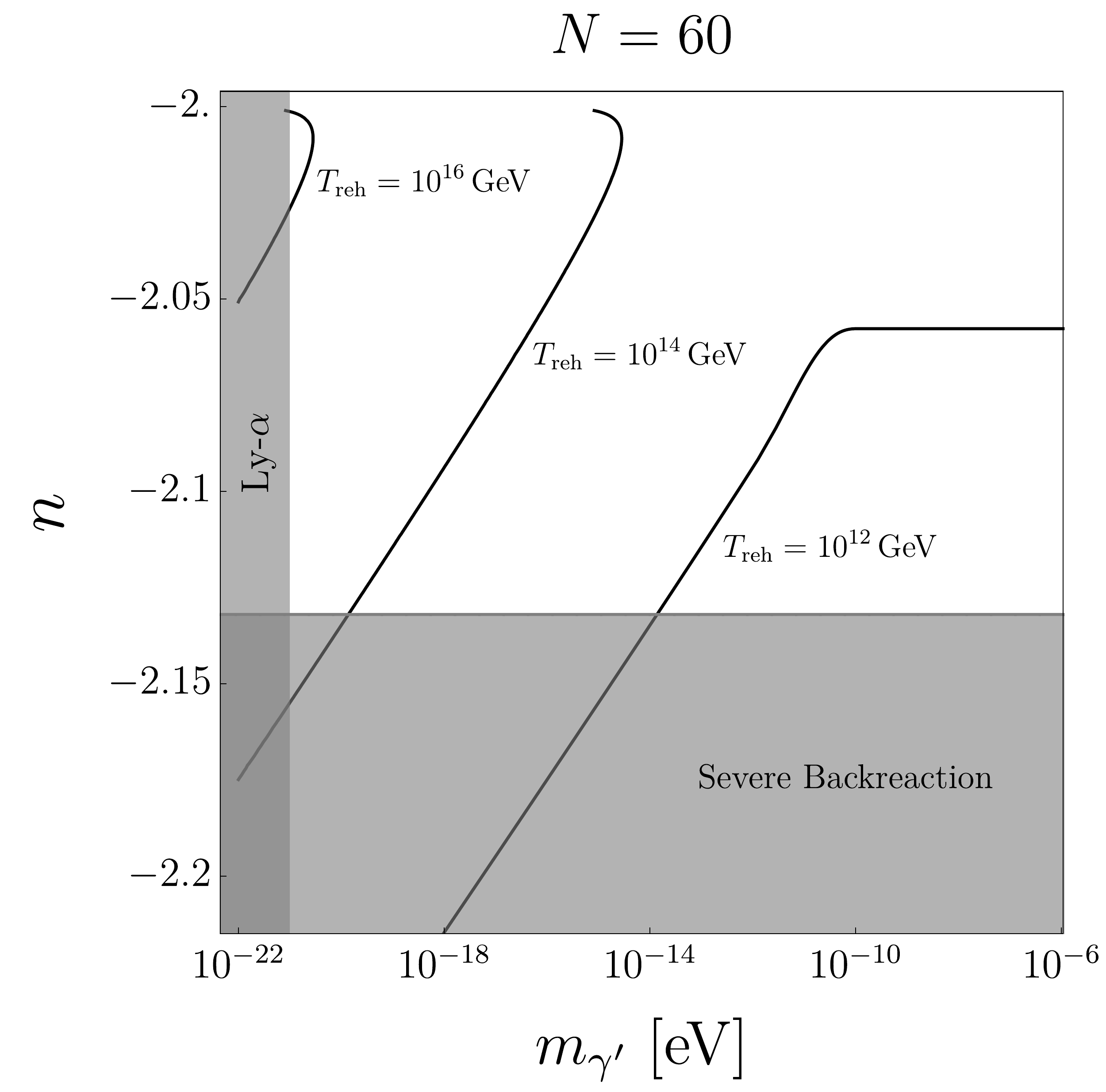}
}
\caption{Plots of parameters that give the correct relic abundance of the dark photon dark matter as a function of $m_{\gamma'}$ and $n$.
The left and right panels correspond to the cases of $N=50$ and $N=60$, respectively.
We here take $g_*(t_0) = 2$ and relate $\rho(t_{\rm end})$ and $T_{\rm reh}$ by \eqref{instant_reh}.
The three solid curves denote the cases where the reheating temperature is $T_{\rm reh} = 10^{16}, 10^{14}, 10^{12} \, \rm GeV$.
The lower gray region is excluded by the constraint to avoid strong backreaction of the produced dark photon onto the background inflationary dynamics. 
The dark photon mass of $m_{\gamma'} \lesssim 10^{-21} \, \rm eV$ is excluded by the constraints from Lyman-$\alpha$ forest \cite{Irsic:2017yje} and black-hole superradiance \cite{Davoudiasl:2019nlo}.}
\label{fig:nDN}
\end{figure}

Figure~\ref{fig:nDN} shows the available parameter values of $m_{\gamma'}$ and $n$ that can give the correct relic abundance of the dark photon dark matter. The conditions that are necessarily respected, i.e.~backreaction constraint \eqref{cond_backreaction}, non-relativistic condition at equality \eqref{cond_nonrel} and masslessness condition during inflation \eqref{cond_masslessness}, are superimposed.
Interestingly enough, the last condition \eqref{cond_masslessness} is extremely severe for the cases $n > 2$, excluding all the parameter regions, and thus we show no plots for $n>2$.
For $n < -2$, on the other hand, the conditions \eqref{cond_nonrel} and \eqref{cond_masslessness} are less stringent than \eqref{cond_backreaction} for the values of the dark photon mass $m_{\gamma'} \lesssim \mu {\rm eV}$, and they even do not show up in the figures.
The main message from Figure~\ref{fig:nDN} is that the correct relic abundance is realized with a dark photon mass extending down to $10^{-21} \, \rm eV$, which is an interesting possibility for a light dark matter.

To draw the plots in Figure~\ref{fig:nDN}, we have set the number of e-folds of inflation, $N=50$ on the left panel and $N=60$ on the right. For other values, we have chosen $h = 0.67$, $g_*(t_{\rm NR}) = g_*(t_0) = 2$, $g_*(t_{\rm reh}) = 106.75$, $\Omega_m h^2 = 0.143$ and $\Omega_{\gamma'} h^2 = \Omega_c h^2 = 0.120$. For simplicity, we also assume an instantaneous reheating right after the inflation, relating the inflationary energy scale and the reheating temperature by 
\begin{equation}
\rho(t_{\rm end}) = \frac{\pi^2}{30} \, g_*(t_{\rm reh}) T_{\rm reh}^4 \; .
\label{instant_reh}
\end{equation}
This last assumption is in general not necessary and can be relaxed. 
For most of the parameter choices, the relic abundance is given by \eqref{relic_relativistic}, but only the part of the rightmost ($T_{\rm reh} = 10^{12} \, {\rm GeV}$) curve on the right panel ($N=60$) is given by \eqref{relic_nonrel}. This can be seen by the flattening of the curve toward the right border -- the abundance does not depend on the dark photon mass, as can be seen in \eqref{relic_nonrel}. This is because the dark photon is non-relativistic already at the end of inflation, and the inflationary production mechanism, as well as the subsequent evolution, is controlled only by the cosmological expansion.

\section{Conclusion}\label{discussion}

In this paper, we have discussed the production of a light dark photon DM
without the assistance of any field other than the inflaton.
Our dark photon directly couples to the inflaton, and thus the model is minimal in the sense that we do not require any additional exotic ingredients.
We considered a coupling of the dark photon field to the inflaton $\varphi$ in the form $I^2(\varphi)FF$,
where $I$ is some function of $\varphi$.
The classical dark photon field is produced due to the motion of $\varphi$, through which the former experiences the ``horizon exit'' of quantum fluctuations during inflation.

This coupling can induce a highly efficient, copious production of long-wavelength modes of the dark photon. Its spectrum is peaked around the wavenumber that exits the horizon at the earliest time and subsequently grows exponentially due to the expansion during inflation. Since the produced dark photon carries small momentum, this mechanism circumvents the requirement for a dark photon to be non-relativistic in the early time and allows a parameter domain with a sub-$\mu {\rm eV}$ DM mass.
For a successful scenario with the correct dark photon DM abundance, a few consistency conditions are imposed: (i) the backreaction of the produced dark photon onto the background inflationary dynamics needs to be negligible, (ii) the dark photon mass should be much smaller than the inflationary Hubble parameter to avoid mass suppression, and (iii) the dark photon necessarily becomes non-relativistic before the matter-radiation equality. 

Interestingly, the condition (ii) excludes the entire parameter region in which $I(\varphi)$ is an increasing function in time.
For the other case where $I(\varphi)$ is decreasing, imposing the requirement (i) automatically satisfies (ii) and (iii).
Yet, with all these constraints respected, we have shown that the correct relic abundance is realized with a dark photon mass extending down to $10^{-21} \, \rm eV$. This is in the realm of fuzzy dark matter, and the mass range of dark matter below this value is excluded by the Lyman-$\alpha$ forest observations and the superradiance constraints from a supermassive black hole.%
\footnote{In this study we assume negligible coupling of the dark photon to any other fields. For potential extra constraints in the cases where the dark photon couples to a $B$ or $B-L$ charge, we refer to refs.~\cite{Graham:2015ifn,Pierce:2018xmy}.}
As can be seen in Figure~\ref{fig:nDN}, the time dependence of the function $I$ should be close to $\sim a^{-2}$, which implies that the dark photon spectrum is close to scale invariant. Although we have assumed a simple relation between $I$ and $a$ to perform analytical calculation, this conclusion itself is expected to hold more generally.

A few issues still remain to be discussed. Firstly, we have not specified the origin of the dark photon mass. If the mass arises from some Higgs-like mechanism in the dark sector, the produced dark photon may backreact on the Higgs potential.
In this case, one may need to solve the equations of motion for the dark photon and the Higgs field simultaneously, because their classical configuration is in general time-dependent in cosmological setups. This approach may also clarify the nature of symmetry breaking/restoration in the dark sector.

On the other hand, we have (somewhat carefully) avoided mentioning the explicit form of $I(\varphi)$ and its UV physics. This is beyond our current scope, but depending on its origin, e.g.~dimensional reduction factor or radiative corrections of the dark $U(1)$ charge, the dark photon mass might also acquire a dependence on $\varphi$. This can potentially modify our calculation of the production and allowed parameter region for DM abundance.

The inflaton typically starts oscillating after the end of inflation until reheating. This oscillation may in principle cause an enhancement of the dark photon field through parametric resonance. However, in our scenario, most of the dark photon modes produced during inflation are well outside the horizon by the end of inflation, and thus the resonance would have to be broad in order to amplify these $k \to 0$ modes. While one may be able to cook up a form of $I(\varphi)$ that can achieve a broad resonance, simple coupling forms typically lead only to a narrow type.

Finally, we have assumed that our dark photon has no mixing with the visible photon. This way, placing only the Lyman-$\alpha$ bound as an observational constraint is self-consistent, but inclusion of the mixing is also a phenomenologically interesting direction to consider.
We would like to keep the present work as a proposal of production mechanism, and leave these intriguing issues to future studies.

\subsection*{Note added}

While this paper was being completed, an overlapping work of ref.~\cite{1792058} appeared,
where CMB constraints due to the produced dark photon are also discussed.
The scenario we consider corresponds to the case (i) for $\gamma < -4$ ($n < -2$ for us) in Section 3 of \cite{1792058}. Our result shows that the open parameter space is relatively narrow, as only the region $-2.15 \lesssim n < -2$ is allowed from Figure \ref{fig:nDN}, which is consistent to the qualitative claim of \cite{1792058}. Yet we have quantitatively shown that there exists an open parameter space in which our scenario is viable. The reason is related to the fact that our dark photon is originated from vacuum fluctuations and the perturbative treatment is always ensured, see footnote \ref{foot:treatment}.
Another side note is that the mass range of dark photon they consider is much heavier than that of our interest.

\section*{Acknowledgements}

We would like to thank Masahito Yamazaki for discussions and Keshav Dasgupta, Peter Denton and Yevgeny Stadnik for useful comments.
Y.N. is grateful to KEK and Kavli IPMU for their hospitality during the COVID-19 outbreak.

\appendix

\section{The longitudinal contribution}\label{longitudinalcontribution}

In this Appendix, we estimate the energy density of the longitudinal mode just after inflation. Using the expression of $z_k$ in \eqref{Xdef}
and the time dependence of $I$ in \eqref{formofI} in de Sitter limit, we find
\begin{equation}
\begin{split}
\frac{\partial_\tau^2 z_k}{z_k} = \frac{1}{\tau^2} \, 
\frac{2 p^4 - \left( 2 n^2 - 7n + 1 \right) p^2 M^2 + n \left( n+1 \right) M^4}{\left( p^2 + M^2 \right)^2} \; ,
\label{zppz}
\end{split}
\end{equation}
where we have defined $p \equiv {k}/{a}$ and $M \equiv {m_{\gamma'}}/{I}$.
We solve the EOM of  \eqref{EOM_X} for each case of $p \gg M$ and $p \ll M$.
In the case of $p \gg M$, the mass term is negligible compared to the momentum term.
Then the EOM of the longitudinal mode, \eqref{EOM_X}, reads
\begin{equation}
\begin{split}
\partial_\tau^2 X_k + \left( k^2 - \frac{2}{\tau^2} \right) X_k \simeq 0 \; , \qquad
p \gg M \; .
\end{split}
\end{equation}
The solution to this equation, with the Bunch-Davies initial condition, is given by
\begin{equation}
X_k \simeq 
= \frac{{\rm e}^{- i k \tau}}{\sqrt{2k}} \left( \frac{1}{- k \tau} - i \right) \; , \qquad
p \gg M \; ,
\label{Xsol_1}
\end{equation}
up to an arbitrary constant phase, where $H_\nu^{(1)}(x)$ is the Hankel function of the first kind.

On the other hand, in the case of $p \ll M$, the EOM reads
\begin{equation}
\partial_\tau^2 X_k - \frac{n \left( n+1 \right)}{\tau^2} \, X_k \simeq 0 \; , \qquad
p \ll M \; .
\end{equation}
Note that this is the same form of the equation as for the transverse modes, \eqref{EOM_V_Dless_2}.
The solution can be written as
\begin{equation}
X_k \simeq 
C_1 \left( - \tau \right)^{n+1} + \frac{C_2}{\left( - \tau \right)^n} \; ,
\qquad p \ll M \; ,
\label{Xsol_2}
\end{equation}
%
where $C_{1,2}$ are integration constants.
We now connect two solutions of \eqref{Xsol_1} and \eqref{Xsol_2} at the conformal time $\tau_{\mathrm{NR},k}$ when $p = M$
by requiring 
\begin{equation}
\begin{split}
&X_k \, \big\vert_{\tau = \tau_{\mathrm{NR},k} , \, p \gg M} = X_k \, \big\vert_{\tau = \tau_{\mathrm{NR},k} , \, p \ll M} \; , \\[2ex]
&z_k \, \partial_\tau \left( \frac{X_k}{z_k} \right) \, \bigg\vert_{\tau = \tau_{\mathrm{NR},k} , \, p \gg M} = z_k \, \partial_\tau \left( \frac{X_k}{z_k} \right) \, \bigg\vert_{\tau = \tau_{\mathrm{NR},k} , \, p \ll M} \; ,
\end{split}
\end{equation}
which amounts to setting the integration constants as
\begin{equation}
\begin{split}
C_1 = \sqrt{\frac{k}{2}} \, \frac{\left( - \tau_{\mathrm{NR},k} \right)^{-n}}{2n + 1} \; , \qquad
C_2 = \frac{\left( - \tau_{\mathrm{NR},k} \right)^{n-1}}{\sqrt{2} \, k^{3/2}} \; .
\end{split}
\end{equation}

Since the solutions of the EOM have been obtained as in \eqref{Xsol_1} and \eqref{Xsol_2},
we now estimate the contribution of the longitudinal mode to the energy density.
It is found that both solutions give the blue spectrum
and the energy density $\rho_{\gamma', L}$ is dominated by the case of $p \gg M$.
By cutting off the UV $k$ mode at the horizon crossing, i.e.~$k_{\rm UV} = {\cal O}(-1/\tau)$, we obtain
\begin{equation}
\begin{split}
\langle \rho_{\gamma',L} \rangle \vert 
\simeq \langle \rho_{\gamma',L} \rangle \vert_{p \gg M}
\sim \frac{H^4}{8\pi^2} \; ,
\label{rhoL_approx}
\end{split}
\end{equation}
where we have used the assumption of $m_{\gamma'} / I \ll H$ during inflation.
Comparing this to the contribution from the transverse modes in \eqref{rho_approx}, it is clear that the total energy density of the dark photon is dominated by the transverse modes by the exponential factor $(- k_{\rm min} \tau_{\rm end})^{-(2 \vert n \vert - 4)} \sim {\rm e}^{\left( 2 \vert n \vert - 4 \right) N}$.

\bibliographystyle{utphys}
\bibliography{bib}

\end{document}